# Suppression and restoration of disorder-induced localization mediated by $\mathcal{PT}$-symmetry breaking

Yaroslav V. Kartashov,[1,2,*] Chao Hang,[3] Vladimir V. Konotop,[4] Victor A. Vysloukh,[5] Guoxiang Huang,[3] and Lluis Torner[1,6]

*Corresponding Author: E-mail: Yaroslav.Kartashov@icfo.es

[1]ICFO-Institut de Ciencies Fotoniques, The Barcelona Institute of Science and Technology, 08860 Castelldefels (Barcelona), Spain
[2]Institute of Spectroscopy, Russian Academy of Sciences, Troitsk, Moscow Region, 142190, Russia
[3]State Key Laboratory of Precision Spectroscopy and Department of Physics, East China Normal University, Shanghai 200062, China
[4]Centro de Física Teórica e Computacional and Departamento de Física, Faculdade de Ciências, Universidade de Lisboa, Campo Grande 2, Edifício C8, Lisboa 1749-016, Portugal
[5]Departamento de Fisica y Matematicas, Universidad de las Americas—Puebla, 72820 Puebla, Mexico
[6]Universitat Politecnica de Catalunya, 08034, Barcelona, Spain

**Abstract:** We uncover that the breaking point of the $\mathcal{PT}$-symmetry in optical waveguide arrays has a dramatic impact on light localization induced by the off-diagonal disorder. Specifically, when the gain/loss control parameter approaches a critical value at which $\mathcal{PT}$-symmetry breaking occurs, a fast growth of the coupling between neighboring waveguides causes diffraction to dominate to an extent that light localization is strongly suppressed and statistically averaged width of the output pattern substantially increases. Beyond the symmetry-breaking point localization is gradually restored, although in this regime the power of localized modes grows upon propagation. The strength of localization monotonically increases with disorder at both, broken and unbroken $\mathcal{PT}$-symmetry.

## 1. Introduction

Anderson localization [1] is a universal wave phenomenon that manifests itself in the complete arrest of ballistic expansion of initially localized excitations in a disordered environment [2,3]. As a pure wave phenomenon based on the interference of waves scattered by a random potential, disorder-induced localization was observed with microwaves [4,5], matter [6,7] and acoustic [8] waves. For optical radiation, it has been studied in strongly scattering materials [9], shallow lattices [10–17], plasmonic arrays [18], and disordered fibers [19]. In the particularly important case of disordered potentials that are constant along the direction of light propagation, one observes so-called transverse Anderson localization [11]. Shallow optical waveguide arrays provide a unique platform to explore localization of light, because they readily allow the experimental implementation of different types of controllable disorder [13-15].

Disorder-induced localization has been mainly studied in conservative systems, as it is usually believed that the onset of localization requires multiple scattering by the *real part* of the potential. The possibility of disorder-induced localization in non-Hermitian quantum mechanics in the presence of complex gauge field and real random potential [20] and in media with gain and losses [21] was suggested long ago, but did not receive rigorous further consideration. Recent experiments indicate that even periodic transverse modulation of losses (i.e., of the *imaginary part* of the potential) notably affects the expansion rate of localized excitations [22]. Localization induced by disorder in the imaginary part of the potential was reported in [23] for a tight-binding model.

Complex potentials $\mathcal{U}(\mathbf{x})$ obeying parity-time $\mathcal{PT}$-symmetry, i.e. $\mathcal{U}(\mathbf{x}) = \mathcal{U}^*(-\mathbf{x})$, host properties of both conservative and dissipative systems, which are controlled by the potential parameters [24]. The spectrum of their eigenmodes can remain real (*unbroken $\mathcal{PT}$-symmetry*) if the imaginary part of the potential does not exceed a threshold.

Above threshold, the spectrum becomes complex and the $\mathcal{PT}$-symmetry is said to be *broken*. The relevance of $\mathcal{PT}$-symmetry in optics [25,26] resulted in observation of $\mathcal{PT}$-symmetry breaking and related phenomena in various guiding structures [27-29]. While regular $\mathcal{PT}$-symmetric structures are well studied, the impact of disorder on light propagation in such structures is not understood, especially around the $\mathcal{PT}$-symmetry breaking threshold. A related question was addressed in [30] where within a tight-binding model for $N$ random $\mathcal{PT}$-symmetric impurities (realizing diagonal disorder), it was shown that the symmetry breaking threshold decreases as $N^{-2}$. In [31] it was conjectured that addition of weak diagonal uniform disorder into nonlinear complex two-dimensional lattice leads to breakup of the $\mathcal{PT}$-symmetry. The same conclusion was drawn for binary disordered arrays [32]. It is also relevant to mention here the results indicating on the existence of localization-delocalization transition reported for Harper (also known as Aubry-Andre) model with a $\mathcal{PT}$-symmetric potential [33], representing an intermediate system between regular periodic and disordered ones.

In this paper we introduce a continuous $\mathcal{PT}$-symmetric array with *off-diagonal* disorder, which consists of waveguides individually respecting $\mathcal{PT}$-symmetry. We study disorder-induced light localization under conditions in which the system undergoes $\mathcal{PT}$-symmetry breaking. To ensure that the unbroken symmetry is *preserved* even close to the breaking threshold, we impose specific symmetry requirements also on the refractive index distribution of the disordered arrays. We find that light localization is suppressed at the symmetry-breaking point, and restored in the broken-symmetry regime. All results are obtained in the frames of the Schrödinger equation with complex potential describing propagation of paraxial light beams in the medium with spatially inhomogeneous gain and losses. Both direct propagation and analysis of eigenmodes supported by disordered complex potentials is used to illustrate specific features of Anderson localization in this system.

## 2. Theoretical model

We consider propagation of light along the $\xi$-axis of a disordered waveguide array created by a simultaneous transverse modulation of the refractive index and the gain/losses parameter, governed by the Schrödinger equation for the dimensionless field amplitude $q$:

$$i\frac{\partial q}{\partial \xi} = -\frac{1}{2}\frac{\partial^2 q}{\partial \eta^2} - R(\eta)q. \tag{1}$$

Here $\eta$ is the dimensionless transverse coordinate, $\eta$ and $\xi$ are normalized to the characteristic transverse scale $x_0$ and diffraction length $L_{\text{dif}} = kx_0^2$, respectively, with $k = 2\pi n/\lambda$ being the wavenumber and $n$ being the background refractive index. The function $R(\eta) = R^*(-\eta) = \sum_k R_0(\eta - \eta_k)$ describes superposition of guiding channels centered at $\eta = \eta_k$, $k \in \mathbb{Z}$. The complex potential of a single channel is given by $R_0(\eta) = p_r R_r(\eta) - ip_i R_i(\eta)$, where $R_r(\eta) = \exp(-\eta^6/a^6)$ and $R_i(\eta) = \eta \exp(-\eta^6/a^6)$. For a non-vanishing $p_i$ the system exhibits linear gain in domains with $R_i(\eta) > 0$ and one-photon absorption in domains with $R_i(\eta) < 0$. The quantity $a$ is the waveguide width; the parameters $p_{r,i} \sim \delta n_{r,i} k^2 x_0^2/n$ are proportional to the real $\delta n_r$ and imaginary $\delta n_i$ parts of the refractive index modulation and characterize the waveguide depth and the amplitude of gain/losses. We study off-diagonal disorder with $\eta_k = (k + \delta\eta_k)d$, where $d$ is the regular waveguide spacing and $\delta\eta_k$ is a relative random shift of the center of the $k$-th waveguide. To ensure $\mathcal{PT}$-symmetry, the random shifts $\delta\eta_k$ are set to be uniformly distributed within $\delta\eta_k \in [-s_d, +s_d]$ for $k > 0$, but at the same time are correlated with shifts at $k < 0$, so that $\delta\eta_k = -\delta\eta_{-k}$. The strength of disorder is controlled by the parameter $s_d$.

The conservative limit $(p_i = 0)$ of the model (1) is known as a "one-dimensional liquid", characterized by randomly distributed equivalent "atoms", and it has received considerable attention in the condensed matter literature (see e.g. [34-36]). In particular, it is known that almost all eigenmodes are localized [34]. Calculation of the localization length in such a model can be performed using scattering matrix of the single-waveguide Schrödinger operator $\hat{L} = -(1/2)d^2/d\eta^2 - R_0(\eta)$, i.e. of the eigenvalue problem $\hat{L}w = -bw$ (here $q = w(\eta)e^{ib\xi}$). Formally, if propagation constant $b$ remains real, the results obtained within "liquid model" can be directly applied to our case.

To gain physical insight into the role of the gain/loss strength in localization, we employ tight-binding approximation and consider the lowest eigenfunctions $\{w_k(\eta), w_k^*(\eta)\}$, where $w_k(\eta) = w_0(\eta - kd)$, and $\{w_0(\eta), w_0^*(\eta)\}$ is a conjugated pair from the bi-orthogonal basis of $\hat{L}$: $\hat{L}w_0(\eta) = -b_0 w_0(\eta)$ and $\hat{L}^\dagger w_0^*(\eta) = -b_0 w_0^*(\eta)$ (the dagger stands for the Hermitian conjugation). Each $w_k(\eta)$ is localized around $\eta = kd$ and $\int w_k(\eta)w_m(\eta)d\eta = 0$ for $k \neq m$. We look for the total field in the form $q = \sum_k q_k(\xi)w_k(\eta)\exp(ib_0\xi)$, where $q_k(\xi)$ stand for $\xi$-dependent mode weights. Substituting this expansion into Eq. (1), multiplying by $w_n(\eta)$, and integrating over $\eta$ one arrives at the discrete equation

$$i\frac{dq_n}{d\xi} = -c_{n,n+1}q_{n+1} - c_{n,n-1}q_{n-1}, \qquad (2)$$

where for $|\delta\eta_k| \ll 1$ the coupling constants are given by $c_{n,n\pm1} = c - \delta c_{n,n\pm1}$ and

$$c = \int w_0(\eta) R_0(\eta) w_{\pm1}(\eta) d\eta / \int w_0^2(\eta) d\eta \qquad (3)$$

The coefficients

$$\delta c_{n,n\pm1} = \left[\int \frac{dw_0(\eta)}{d\eta} R_0(\eta) w_{\pm1}(\eta) d\eta / \int w_0^2(\eta) d\eta\right](\delta\eta_n - \delta\eta_{n\pm1})d \qquad (4)$$

describe off-diagonal disorder. If the $\mathcal{PT}$-symmetry in each individual potential well is unbroken, i.e. $b_0$ is real, one can choose a $\mathcal{PT}$-symmetric basis $w_k(\eta) = w_k^*(-\eta)$ that yields *real* coupling constants $c$. If random waveguide shifts are correlated, what in our case means $\delta\eta_k = -\delta\eta_{-k}$, one can show that the random matrix $\delta c_{n,m}$ describing perturbations of coupling constants $c$ is $\mathcal{PT}$-symmetric, i.e. $\delta c_{n,n\pm1} = \delta c_{-n,-n\mp1}^*$ (in our case the operator $\mathcal{P}$ is given by the anti-diagonal matrix with 1 on the anti-diagonal, i.e. with the entries $\mathcal{P}_{nm} = \delta_{n,-m}$, and $\mathcal{T}$ is a complex conjugation). Therefore the total matrix $c_{n,n\pm1} = c - \delta c_{n,n\pm1}$ is a sum of an unperturbed real diagonal matrix and a small $\mathcal{PT}$-symmetric matrix with off-diagonal elements $\delta c_{n,n\pm1}$. For an array with a finite number of waveguides, such perturbation does not lead to complex eigenvalues if $|\delta c_{n,n\pm1}| \ll |c|$ [37].

In order to calculate localization length within the framework of the tight-binding model (2) one looks for a solution of Eq. (2) in the form $q_n \sim e^{i(\beta+c)\xi}\tilde{q}_n$, where $\tilde{q}_n$ are field amplitudes on different sites, that can be found from $\beta\tilde{q}_n = \delta c_{n,n+1}\tilde{q}_{n+1} + \delta c_{n,n-1}\tilde{q}_{n-1}$. This last system was studied in [36] for real finite coupling coefficients $\delta c_{n,m}$ and it was found that $\beta = 0$ corresponds to a delocalized mode. Since for $\beta = 0$ we have $\tilde{q}_{n+1}/\tilde{q}_{n-1} = -\delta c_{n,n-1}/\delta c_{n,n+1}$, the localization length $\ell$ defined by the limit $\ell^{-1} = -\lim(1/2n)\ln|\tilde{q}_n/\tilde{q}_{-n}|$ should *diverge* at $\beta \to 0$ according to the arguments of [36] based on the central-limit theorem, that are fully applicable in our case. This allows us to predict that at the $\mathcal{PT}$-symmetry breaking point for the individual waveguide, i.e. at $p_i = p_i^{sb}$, one should observe *localization-delocalization transition*. Indeed, at the exceptional point $p_i = p_i^{sb}$ the eigenfuctions $w_k(\eta)$ become self-orthogonal [38], i.e. $\int w_k^2(\eta)d\eta = 0$, and coupling constant (3) and its perturbation (4) diverge, while the ratio $\delta\tilde{c}_{n,n\pm1} = \delta c_{n,n\pm1}/c$ remains finite. By defining $\tilde{\beta} = \beta/c$, one can cast tight-binding model in the form $\tilde{\beta}\tilde{q}_n = \delta\tilde{c}_{n,n+1}\tilde{q}_{n+1} + \delta\tilde{c}_{n,n-1}\tilde{q}_{n-1}$, where $\tilde{\beta} \to 0$ and $\delta\tilde{c}_{n,n\pm1}$ remain finite in the exceptional point, thereby supporting conclusion about delocalization of modes. Physically this means that drastic increase of the effective coupling between sites stimulates almost unconstrained expansion of excitations, thereby leading to delocalization. At $p_i > p_i^{sb}$ the $\mathcal{PT}$-symmetry of individual waveguides gets broken and $c_{n,n\pm1}$ coefficients become complex, so that one has to resort to numerical integration to study localization dynamics.

## 3. Results and discussion

To calculate variation of the coupling constant $c$ with $p_i$ we use original model (1) where we set $p_r = 8$ (that for $\lambda = 632$ nm, refractive index $n = 1.45$, and $x_0 = 10$ $\mu$m corresponds to a refractive index modulation depth $\delta n_r \sim 5.6 \times 10^{-4}$), the period of the array $d = 2$ (corresponding to spacing of 20 $\mu$m), and the waveguide width $a = 0.5$ (5 $\mu$m). Such arrays (with controllable off-diagonal disorder) may be realized experimentally by direct fs-laser writing that allows precise control of separation between waveguides [15,16]. Losses were already introduced in such systems using small-scale longitudinal oscillations of waveguide centers [22]. Inhomogeneous gain landscapes can be created in such or similar systems by using inhomogeneous doping with active centers.

In Figure 1(a) we show the dependencies of real $c_r = \text{Re}\, c$ and imaginary $c_i = \text{Im}\, c$ parts of the coupling constant on the gain/loss strength $p_i$ in a *regular* array. The coupling constant remains real below the $\mathcal{PT}$-symmetry breaking point $p_i^{sb} \approx 18.5$ for a single guide. In such regime light experiences standard discrete diffraction. Coupling constant rapidly grows with increasing $p_i$ and diverges as $p_i \to p_i^{sb}$, suggesting that a *drastic enhancement of diffraction* should occur around the symmetry breaking point. Beyond the symmetry breaking point, $c$ acquires a large imaginary part that *decreases* when one goes deeper into the domain with broken symmetry. At the same time, $|c_r|$ remains finite and slowly decreases with $p_i$. Notice that imaginary part of coupling constant does affect the expansion rate of localized excitations. This is obvious from Eq. (2) that for the case of a two-channel system predicts that the amplitudes $q_{1,2}$ evolve as $q_{1,2} = \alpha e^{+(ic_r - c_i)\xi} \pm \beta e^{-(ic_r - c_i)\xi}$, where $\alpha, \beta$ are constants. When $c_i \neq 0$ this expression describes power exchange between channels with period $\pi/c_r$ accompanied by growth of *both* amplitudes $q_{1,2}$ with distance ($\sim e^{c_i\xi}$ at $\xi \to \infty$). Interestingly, there exists a specific point $p_i \approx 30.7$ in the broken-symmetry region where $c_i = 0$. Although in this region the power still grows driven by complex propagation constants $b$, the rate of growth is identical for all channels, so that if the power $U = \int |q|^2 d\eta$ is renormalized at each distance $\xi$, one observes a discrete diffraction pattern very similar to that of a conservative array.

Figure 1(b) depicts the integral width of the output intensity distribution in a *regular* array (introduced as $w_\chi = 1/\chi$, where $\chi = U^{-2} \int |q|^4 d\eta$) versus $p_i$ after propagation up to the distance $\xi = 150$ in the case of single-site excitation. This dependence was calculated using continuous model (1). Rapid growth of the width around the $\mathcal{PT}$-symmetry breaking point (in the periodic array $p_i^{sb} \approx 16.9$) is consistent with behavior of coupling constant in the tight-binding description [Figure 1(a)]. Also, the integral width exhibits a local maximum around the point where $c_i \to 0$.

Enhancement of the coupling strength at $p_i = p_i^{sb}$ suggests that Anderson localization induced by the weak off-diagonal disorder may be *substantially weakened* there. In terms of the mode structure, off-diagonal disorder results in the formation of local defects in the effective refractive index that support localized modes. When the coupling between waveguides grows, the modes broaden to accommodate to the increased strength of the effective diffraction. To study this effect, for each value of $p_i$ we generated $Q = 10^4$ realizations of *disordered* $\mathcal{PT}$-symmetric arrays, each containing $N = 200$ waveguides, and integrated Eq. (1) for each realization using single-site excitation up to a large distance ($\xi = 300$), to ensure that the expansion of the wavepacket is arrested. We calculated the statistically averaged width $w_\chi^{av} = Q^{-1} \sum_{k=1,Q} 1/\chi_k$ and averaged intensity distribution $I_{av} = Q^{-1} \sum_{k=1,Q} |q_k|^2$.

The dependence of the averaged width of the output pattern on $p_i$ for a fixed disorder level ($s_d = 0.1$), shown in Figure 1(c), is the key result of this paper. One observes that $w_\chi^{av}$ rapidly grows when $p_i \to p_i^{sb}$ and acquires a maximal (but finite) value at $p_i = p_i^{sb}$. Therefore, around the symmetry-breaking point, disorder-induced localization is *suppressed*. Subsequent increase of $p_i$ inside the regime of broken $\mathcal{PT}$ symmetry leads to gradual *restoration* of the localization, in accordance with the decrease of the modulus of the complex coupling constant. The result is a strongly non-monotonic $w_\chi^{av}(p_i)$ dependence. Despite the exponential power growth at $p_i > p_i^{sb}$ the output patterns in each realization of disorder remain *spatially localized*, i.e. localization still occurs even in the system with broken symmetry, although now it is achieved not exclusively due to guiding defects, but also due to gain and loss.

Figure 2 shows illustrative examples of light dynamics and formation of averaged intensity distributions in the disordered $\mathcal{PT}$-symmetric arrays. At $p_i < p_i^{sb}$, after an initial stage of expansion, the averaged pattern approaches a steady-state [Figure 2(a) and 2(b)]. Notice, that in all depicted cases this steady-state regime is achieved at distances below maximal propagation distance $\xi = 300$ considered here. For $p_i$ values closer to the $\mathcal{PT}$-symmetry-breaking point, one observes broader patterns [compare Fig. 2(a) and 2(b)]. In the presence of disorder, symmetry-breaking in some array realizations may occur for $p_i$ values slightly below the nominal threshold $p_i^{sb}$ defined for a regular array. However, such deviations in threshold are *small* for the off-diagonal symmetric disorder considered here (naturally, they grow with $s_d$). Thus, if $p_i$ is not too close to the symmetry breaking threshold $p_i^{sb}$ in regular array, the propagation dynamics of the beam in *each realization* of the disordered $\mathcal{PT}$-symmetric array closely resembles that in a usual conservative disordered structure, i.e. one observes intensity oscillations between several channels around the excitation point *without* net amplification or attenuation of the beam [a representative example of such dynamics for one realization of $\mathcal{PT}$-symmetric disorder is shown in Fig. 2(c)]. In this regime, the localization dynamics in a disordered $\mathcal{PT}$-symmetric system is qualitatively similar to that observed in a conservative disordered array. It should be properly appreciated that one can *closely* approach the symmetry breaking threshold $p_i^{sb} \approx 16.9$ and observe localization dynamics similar to that occurring in conservative array only because random waveguide shifts $\delta\eta_k = -\delta\eta_{-k}$ are correlated in the model of $\mathcal{PT}$-symmetric disorder [that, in particular, results in $\mathcal{PT}$-symmetric matrix $c_{n,n\pm1}$ in the tight-binding approximation (2)] and do not lead to pronounced variation of symmetry-breaking threshold $p_i^{sb}$, as long as $s_d$ is small.

However, for sufficiently small $\delta c_{n,n\pm1}$ the reality of the spectrum of the matrix of coupling coefficients persists [37] independently of the fact whether $\mathcal{PT}$-symmetry is violated or not, i.e. the spectrum may remain real even in the case of completely uncorrelated disorder, when $\delta c_{n,n\pm1} \neq \delta c_{-n,-n\mp1}^*$. Thus, if uncorrelated disorder is sufficiently weak, one can expect qualitatively similar (to the $\mathcal{PT}$-symmetric case) behavior of the light localization. The main difference is expected to be quantitative, because now the symmetry breaking should occur at much smaller $p_i$ values. Indeed, we found that if off-diagonal disorder is uncorrelated, for small $s_d \sim 0.1$ one can still observe disorder-induced localization of light without net attenuation or amplification in each realization of the array, but only for $p_i$ values substantially smaller than the symmetry breaking threshold $p_i^{sb}$ in regular array [see blue dots in Fig. 1(c) obtained for disorder with uncorrelated shifts of waveguides illustrating that statistically averaged beam width $w_\chi^{av}$ also grows with $p_i$ in this case]. For uncorrelated disorder with $s_d = 0.1$ the $\mathcal{PT}$ symmetry in some realizations gets broken already for $p_i > 12$, resulting in beam amplification/attenuation, so we do not show $w_\chi^{av}$ for larger $p_i$ range in this case.

In the model with $\mathcal{PT}$-symmetric disorder and in the regime $p_i < p_i^{sb}$, where $\mathcal{PT}$-symmetry remains unbroken for a given disorder level $s_d$, the onset of disorder-induced localization can be clearly inferred from statistically averaged output intensity distributions depicted in Fig. 3(a) that have triangular shapes in the logarithmic scale used in the figure. Notice, that averaged output intensity distribution notably broadens when one approaches the symmetry-breaking point. These patterns are qualitatively similar to those encountered in conservative disordered waveguide arrays. It should be stressed that unbroken $\mathcal{PT}$-symmetry in absolutely all realizations of disordered array is a necessary condition for observation of averaged output patterns with exponential tails.

To confirm that in the regime with unbroken $\mathcal{PT}$-symmetry the width of statistically averaged intensity distribution does not change even on large propagation distances, once steady-state regime is reached, and to show that the results are not affected by the finite-size effects, in Fig. 4(a) we show the evolution of statistically averaged beam width $w_\chi^{\mathrm{av}}$ up to doubled propagation distance $\xi=600$ in the arrays containing two times more waveguides ($N=400$) than arrays used above. We show $w_\chi^{\mathrm{av}}(\xi)$ at $p_\mathrm{i}=10$ for two disorder levels $s_\mathrm{d}=0.10$ and $s_\mathrm{d}=0.15$ that guarantees that $\mathcal{PT}$-symmetry is unbroken in each realization. In this case one observes that averaged width of the pattern only slightly oscillates in the steady-state regime (the amplitude of oscillations decreases with disorder level and number of realizations over which averaging is performed). Averaged transverse intensity distributions at distances $\xi=300$ and $\xi=600$ are *nearly indistinguishable*, indicating on the fact that expansion of the wavepacket is completely arrested by disorder [Fig. 4(b)]. Moreover, increase of the system size does not lead to any noticeable changes in the averaged patterns, as seen from comparison of solid and dotted curves in Fig. 4(b).

Somewhat different situation is encountered in the regime with $p_\mathrm{i} \geq p_\mathrm{i}^{\mathrm{sb}}$, where $\mathcal{PT}$-symmetry is broken, even if disorder is symmetric. In this case the power of propagating beam *always increases*, since it unavoidably excites exponentially growing modes. Representative dynamics for one disorder realization for this case is illustrated in Figure 2(d), where we renormalized the power at each distance $\xi$ for convenience. The mode experiencing the highest net gain eventually dominates over all other modes, but the main outcome of our simulations is that *in each particular realization* the output pattern remains *localized* despite overall exponential power growth, as confirmed by Fig. 2(d). Because locations of dominant modes with highest gain vary in different realizations, in this regime exponential tails in *averaged* intensity distributions completely disappear (with the only exception, observed around the point $p_\mathrm{i} \sim 35$), even if power is renormalized in each realization. However, because the width $w_\chi$ of the output pattern always remains *finite* even in this regime, one can still talk about *disorder-induced localization* and determine the averaged output width depicted in Fig. 1(c) even for the $p_\mathrm{i} > p_\mathrm{i}^{\mathrm{sb}}$ case. Interestingly, in the region $p_\mathrm{i} \sim 35$, which is close to the point where $c_\mathrm{i} \to 0$, one again observes rich transverse dynamics in individual realizations [compare Fig. 2(e) with Fig. 2(d)]. It is only around this *unique* point in the broken-symmetry domain, where most of localized modes of disordered system acquire very close imaginary parts of eigenvalues, one can still encounter exponential tails in averaged intensity distributions [see Figs. 2(f) and 3(b)].

Increasing level of disorder always results in localization enhancement, under conditions of both, broken and unbroken $\mathcal{PT}$-symmetry. This behavior is summarized in Fig. 3(c), where the averaged output width monotonically decreases with $s_\mathrm{d}$ at $p_\mathrm{i} < p_\mathrm{i}^{\mathrm{sb}}$ and at $p_\mathrm{i} > p_\mathrm{i}^{\mathrm{sb}}$.

The above results, based on direct integration of Eq. (1), are confirmed by the statistical analysis of the eigenmodes $q=w(\eta)\exp(ib\xi)$ of the disordered arrays. Examples of strongly localized modes associated to the propagation constants $b$ with maximal real part are depicted in Figure 5. From the entire spectrum of guided modes of the random potential, such modes usually are the narrowest. They typically form in two guides with the minimal separation between their centers, and are characterized by strong internal currents from the domains featuring gain into the domains featuring losses. The internal mode structure is different below and above the symmetry-breaking point, consistent with the growing role that the internal energy flows play in the localization process. If disorder is symmetric, the field modulus distribution of the eigenmodes is symmetric too. The whole array consists of $N$ waveguides, and its highest-order mode with index $N$ exhibits the smallest $\operatorname{Re}b$ and is the most extended. At a given disorder level $s_\mathrm{d}$, the symmetry-breaking caused by the growth of $p_\mathrm{i}$ occurs first for such mode around the value $p_\mathrm{i}^{\mathrm{sb}} \approx 16.9$. Symmetry-breaking for modes $N-1$, $N-2$,... occurs for progressively larger values of $p_\mathrm{i}$, so that the lowest-order (narrowest) mode acquires a complex propagation constant at the largest value of $p_\mathrm{i} \approx 21.8$.

The dependence of the statistically averaged width and propagation constant on $p_\mathrm{i}$ for modes with $b_\mathrm{m}=\max(\operatorname{Re}b)$ is depicted in Fig. 6(a). The dependence $w_\chi^{\mathrm{av}}(p_\mathrm{i})$ for the $b_\mathrm{m}$ mode *closely resembles* that encountered by the direct integration of Eq. (1) [Fig. 1(c)]. This further confirms the *strong impact* of the gain/loss strength on the strength of disorder-induced localization. Fig. 6(b) shows the evolution of the averaged width and gain for modes with $g_\mathrm{m} = \max(\mathrm{Im}\, b)$ that grow most rapidly, when the symmetry is broken. The averaged width of the most rapidly growing mode (defined for $p_\mathrm{i} > p_\mathrm{i}^\mathrm{sb}$) decreases with $p_\mathrm{i}$, a result that is consistent with Fig. 1(c). The standard deviations of $g_\mathrm{m}$ and $b_\mathrm{m}$ acquire their maximal values very close to the symmetry-breaking points of the corresponding modes [Fig. 6(c)]. Finally, the simultaneous localization enhancement for the mentioned modes with increasing disorder is confirmed by Fig. 6(d).

## 4. Conclusions and outlook

In conclusion, we predicted that breakup of $\mathcal{PT}$-symmetry in systems with off-diagonal disorder suppresses localization observable when the symmetry is unbroken. The light localization is restored beyond the symmetry-breaking point and becomes stronger again deeply inside the regime of broken symmetry. The implications of this result are potentially far-reaching. From a fundamental standpoint, it shows that the imaginary part of the guiding potential may be responsible for localization-delocalization transition that can now be observed directly due to possibility to tune parameters of the active system *in situ*. From the point of view of applications, it opens interesting possibilities for controlling light localization and evolution dynamics by manipulating gain and losses landscapes. Similarities of mathematical models utilized in paraxial optics, quantum mechanics, and meanfield theory of Bose-Einstein condensation, as well as already reported realizations of $\mathcal{PT}$-symmetric structures utilizing plasmonic waveguides, cavity polaritons, bulk metamaterials and metasurfaces, suggest that the phenomena reported here can be observed in a much wider class of physical systems. In particular, the extension to acoustics and cavity networks is anticipated, since these systems allow introduction of controllable disorder combined with complex potentials.

After submission of this paper we were informed about alternative suggestion for suppression of Anderson localization in the presence of disorder [39], which is based on the robust one-way transport in non-Hermitian lattices.

**Acknowledgements**: The work of VVK was supported by the FCT grants (Portugal) UID/FIS/00618/2013 and PTDC/FIS-OPT/1918/2012. The work of CH and GH was supported by NSFC under Grants No. 11474099 and No. 11475063. YVK and LT were supported by Fundacio Privada Cellex and the Severo Ochoa Excellence program.

## References

[1] P. W. Anderson, Phys. Rev. **109**, 1492 (1958). E. Abrahams, P. W. Anderson, D. C. Licciardello, and T. V. Ramakrishnan, Phys. Rev. Lett. **42**, 673 (1979).
[2] T. Brandes and S. Kettemann, *The Anderson Transition and its Ramifications --- Localization, Quantum Interference, and Interactions* (Berlin: Springer Verlag, 2003).
[3] M. Segev, Y. Silberberg, and D. N. Christodoulides, Nat. Photon. **7**, 197 (2013); D. S. Wiersma, Nat. Photon. **7**, 188 (2013).
[4] R. Dalichaouch, J. P. Armstrong, S. Schulz, P. M. Platzman, and S. L. McCal, Nature **354**, 53 (1991).
[5] P. Pradhan and S. Sridar, Phys. Rev. Lett. **85**, 2360 (2000).
[6] J. Billy, V. Josse, Z. C. Zuo, A. Bernard, B. Hambrecht, P. Lugan, C. Clement, L. Sanchez-Palencia, P. Bouyer, and A. Aspect, Nature **453**, 891 (2008).
[7] G. Roati, C. D. Errico, L. Fallani, M. Fattori, C. Fort, M. Zaccanti, G. Modugno, M. Modugno, and M. Inguscio, Nature **453**, 895 (2008).
[8] H. Hu, A. Strybulevych, J. H. Page, S. E. Skipetrov, and B. A. van Tiggelen, Nature Phys. **4**, 945 (2008).
[9] S. John, Phys. Rev. Lett. **53**, 2169 (1984); D. S. Wiersma, P. Bartolini, A. Legendijk, and R. Righini, Nature **390**, 671 (1997).
[10] S. S. Abdullaev and F. K. Abdullaev, Sov. J. Radiofizika **23**, 766 (1980).
[11] H. de Raedt, A. Lagendijk, and P. de Vries, Phys. Rev. Lett. **62**, 47 (1989).
[12] T. Pertsch, U. Peschel, J. Kobelke, K. Schuster, H. Bartelt, S. Nolte, A. Tünnermann, and F. Lederer, Phys. Rev. Lett. **93**, 053901 (2004).

[13] T. Schwartz, G. Bartal, S. Fishman, and M. Segev, Nature **446**, 52 (2007).
[14] Y. Lahini, A. Avidan, F. Pozzi, M. Sorel, R. Morandotti, D. N. Christodoulides, and Y. Silberberg, Phys. Rev. Lett. **100**, 013906 (2008); G. Kopidakis, S. Komineas, S. Flach, and S. Aubry, Phys. Rev. Lett. **100**, 084103 (2008).
[15] A. Szameit, Y. V. Kartashov, P. Zeil, F. Dreisow, M. Heinrich, R. Keil, S. Nolte, A. Tünnermann, V. A. Vysloukh, and L. Torner, Opt. Lett. **35**, 1172 (2010); M. I. Molina, N. Lazarides, and G. P. Tsironis, Phys. Rev. E **85**, 017601 (2012); D. M. Jovic, Y. S. Kivshar, C. Denz, and M. R. Belic, Phys. Rev. A **83**, 033813 (2011).
[16] D. M. Jovic, M. R. Belic, and C. Denz, Phys. Rev. A **84**, 043811 (2011); S. Stützer, Y. V. Kartashov, V. A. Vysloukh, A. Tünnermann, S. Nolte, M. Lewenstein, L. Torner, and A. Szameit, Opt. Lett. **38**, 1488 (2013).
[17] V. E. Lobanov, Y. V. Kartashov, V. A. Vysloukh, and L. Torner, Phys. Rev. A **88**, 053829 (2013).
[18] X. Shi, X. Chen, B. A. Malomed, N. C. Panoiu, and F. Ye, Phys. Rev. B **89**, 195428 (2014).
[19] M. Leonetti, S. Karbasi, A. Mafi, and C. Conti, Phys. Rev. Lett. **112**, 193902 (2014); S. Karbasi, R. J. Frazier, K. W. Koch, T. Hawkins, J. Ballato, and A. Mafi, Nat. Commun. **5**, 3362 (2014).
[20] N. Hatano and D. R. Nelson, Phys. Rev. Lett. **77**, 570 (1996).
[21] A. A. Asatryan, N. A. Nicorovici, L. C. Botten, C. Martijn de Sterke, P. A. Robinson, and R. C. McPhedran, Phys. Rev. B **57**, 13535 (1998).
[22] T. Eichelkraut, R. Heilmann, S. Weimann, S. Stützer, F. Dreisow, D. N. Christodoulides, S. Nolte and A. Szameit, Nat. Commun. **4**, 3533 (2013).
[23] A. Basiri, Y. Bromberg, A. Yamilov, H. Cao, and T. Kottos, Phys. Rev. A **90**, 043815 (2014).
[24] C. M. Bender, Rep. Prog. Phys. **70**, 947 (2007).
[25] A. Ruschhaupt, F. Delgado, and J. G. Muga, J. Phys. A: Math. Gen. **38**, L171 (2005).
[26] R. El-Ganainy, K. G. Makris, D. N. Christodoulides, and Z. H. Musslimani, Opt. Lett. **32**, 2632 (2007).
[27] A. Guo, G. J. Salamo, D. Duchesne, R. Morandotti, M. Volatier-Ravat, and V. Aimez, G. A. Siviloglou, and D. N. Christodoulides, Phys. Rev. Lett. **103**, 093902(2009); C. E. Rüter, K. G. Makris, R. El-Ganainy, D. N. Christodoulides, M. Segev, and D. Kip, Nat. Phys. **6**, 192 (2010).
[28] K. G. Makris, R. El-Ganainy, D. N. Christodoulides, and Z. H. Musslimani, Phys. Rev. Lett. **100**, 103904 (2008).
[29] S. Longhi, Phys. Rev. Lett. **103**, 123601 (2009).
[30] O. Bendix, R. Fleischmann, T. Kottos, and B. Shapiro, Phys. Rev. Lett. **103**, 030402 (2009).
[31] D. M. Jovic, C. Denz, and M. R. Belic, Opt. Lett. **37**, 4455 (2012).
[32] C. Mejia-Cortes and M. I. Molina, Phys. Rev. A **91**, 033815 (2015).
[33] A. Jazaeri and I. I. Satija, Phys. Rev. E **63**, 036222 (2001); C. H. Liang, D. D. Scott, and Y. N. Joglekar, Phys. Rev. A **89**, 030102(R) (2014); C. Yuce, Phys. Lett. A **378**, 2024 (2014).
[34] B. Y. Tong, Phys. Rev. A **1**, 52 (1970).
[35] K. Ishii, Suppl. Prog. Theor. Phys, Nº 53, 77 (1973).
[36] G. Theodorou and M. H. Cohen, Phys. Rev. B **13**, 4597 (1976).
[37] E. Caliceti, S. Graffi, and J. Sjöstrand, J. Phys. A: Math. Gen. **38**, 185 (2005); D. E. Pelinovsky, D. A. Zezyulin, and V. V. Konotop, J. Phys. A: Math. Theor. **47**, 085204 (2014).
[38] N. Moiseyev, *Non-Hermitian quantum mechanics* (Cambridge: Cambridge University Press, 2009).
[39] S. Longhi, D. Gatti, and G. Della Valle, "Robust light transport in non-Hermitian photonic lattices", arXiv:1503.08787 (2015).

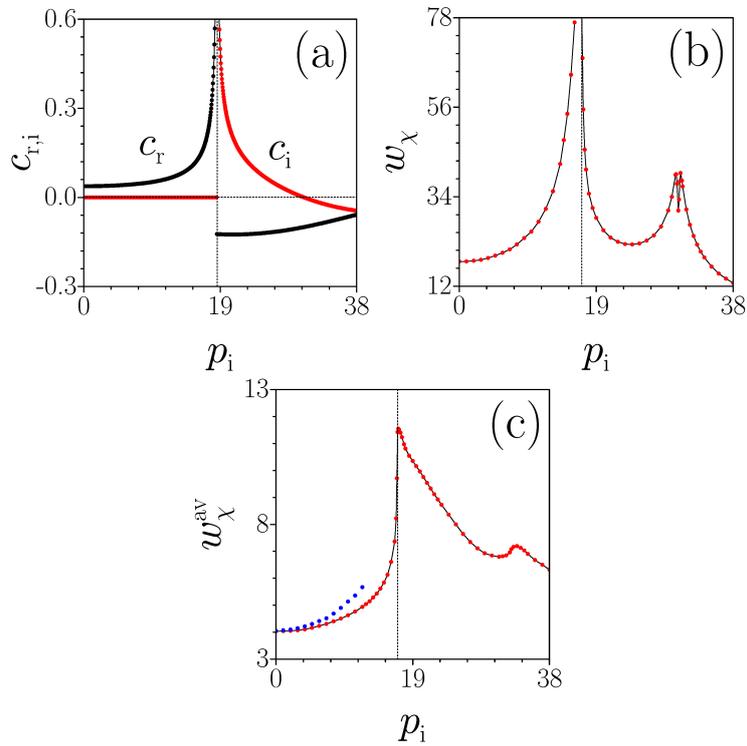

**Figure 1.** (a) Real $c_r$ and imaginary $c_i$ parts of the coupling constant versus $p_i$. The vertical dashed line stands for the $\mathcal{PT}$-symmetry-breaking point in an isolated waveguide. (b) Output beam width at $\xi=150$ versus $p_i$ in a regular $\mathcal{PT}$-symmetric array. (c) Red dots - statistically averaged output width at $\xi=300$ versus $p_i$ in the $\mathcal{PT}$-symmetric waveguide array with $s_d=0.1$. Blue dots – averaged width versus $p_i$ in the array with uncorrelated disorder at $s_d=0.1$. The dashed lines in (b),(c) stand for the symmetry breaking point in regular array. Here and in all plots $p_r=8$.

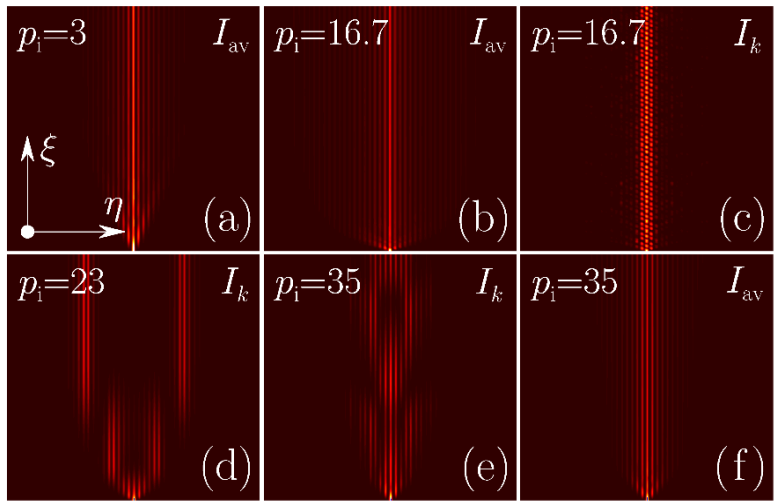

**Figure 2.** Statistically averaged intensity distributions (a),(b) and propagation dynamics for illustrative realizations (c) in disordered array with $p_\text{i} < p_\text{i}^\text{sb}$. Examples of propagation dynamics in two particular realizations (d),(e) and statistically averaged intensity distribution (f) in arrays with $p_\text{i} > p_\text{i}^\text{sb}$. In all cases the propagation distance is $\xi = 300$ and $s_\text{d} = 0.1$.

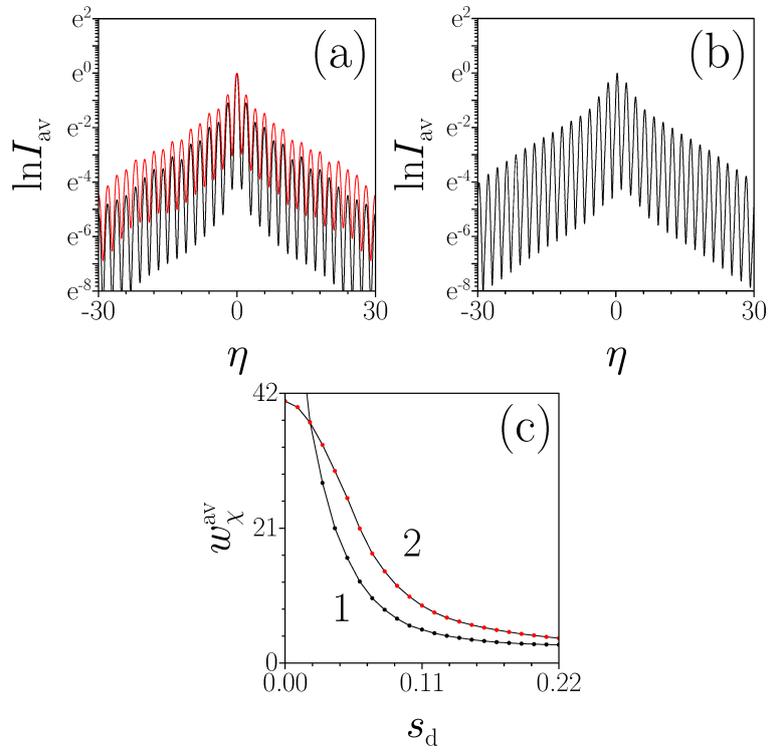

**Figure 3.** Averaged output intensity distributions at $p_i = 2$ [panel (a), black line], $p_i = 16.7$ [panel (a), red line], and $p_i = 35$ [panel (b)] for $s_d = 0.1$. (c) Averaged output width versus $s_d$ at $p_i = 15 < p_i^{sb}$ (curve 1) and $p_i = 19 > p_i^{sb}$ (curve 2).

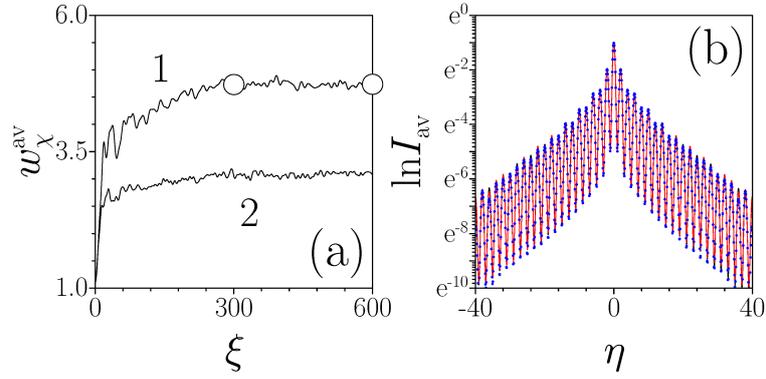

**Figure 4.** (a) Statistically averaged beam width versus propagation distance at $p_i=10$ for $s_d=0.10$ (curve 1) and $s_d=0.15$ (curve 2) in the array containing $N=400$ waveguides. Circles correspond to averaged intensity distributions at $\xi=300$ and $\xi=600$, shown in panel (b) with black and red lines, respectively. Blue dots in (b) show averaged intensity distribution at $\xi=300$ calculated for the array with $N=200$ waveguides. In all cases disorder is $\mathcal{PT}$-symmetric.

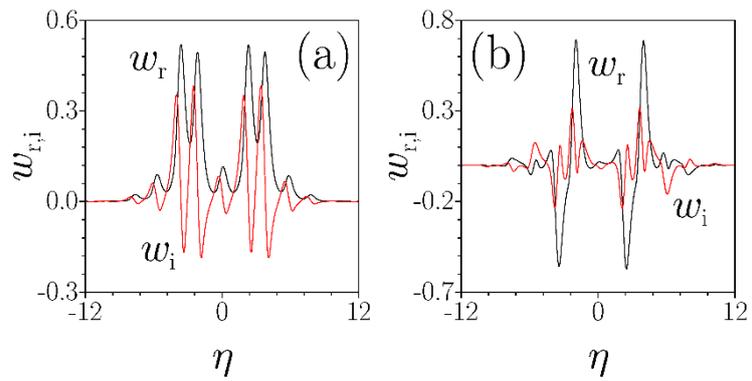

**Figure 5.** Shapes of localized eigenmodes of a disordered array corresponding to the eigenvalues with maximal real part, for $p_i = 19$ (a) and $p_i = 40$ (b). In both cases $s_d = 0.15$.

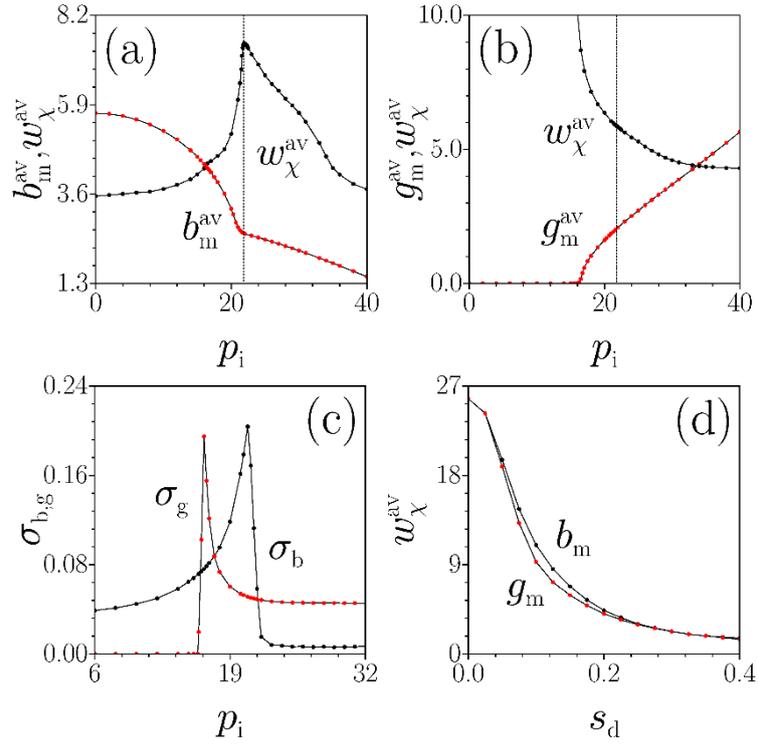

**Figure 6.** (a) Averaged width and propagation constant for the mode corresponding to the eigenvalue with maximal real part, versus $p_i$ at $s_d = 0.15$. (b) Averaged width and gain for the mode corresponding to the eigenvalue with maximal imaginary part, versus $p_i$ at $s_d = 0.15$. Dashed lines in (a),(b) correspond to the symmetry-breaking point for the mode with $b = b_m$. (c) Standard deviations for $b_m^{av}$ and $g_m^{av}$ versus $p_i$ at $s_d = 0.15$. (d) Averaged width of modes with maximal real (curve $b_m$) and maximal imaginary (curve $g_m$) parts of the eigenvalues, versus $s_d$ at $p_i = 21.5$.